\RequirePackage{fixltx2e}
\documentclass[aip,jcp,reprint,floatfix]{revtex4-1}
\usepackage{graphicx} 
\usepackage{dcolumn}
\usepackage{amsmath,amsfonts} %
\usepackage{braket} %

\newcommand{\rr}{\mathbf{r}} %
\newcommand{\RR}{\mathbf{R}} %
\newcommand{\qq}{\mathbf{q}} %
\newcommand{\pp}{\mathbf{p}} %
\newcommand{\eq}[1]{Eq.~(\ref{#1})} %
\newcommand{\eqs}[1]{Eqs.~(\ref{#1})} %
\newcommand{\fig}[1]{Fig.~\ref{#1}} %
\newcommand{\bea}{\begin{eqnarray}}
\newcommand{\eea}{\end{eqnarray}}

\begin{document}

\title{Topologically correct quantum nonadiabatic formalism for on-the-fly dynamics}

\author{Lo{\"i}c Joubert-Doriol} %
\affiliation{Department of Physical and Environmental Sciences,
  University of Toronto Scarborough, Toronto, Ontario, M1C 1A4,
  Canada} %
\affiliation{Chemical Physics Theory Group, Department of Chemistry,
  University of Toronto, Toronto, Ontario M5S 3H6, Canada} %
\author{Janakan Sivasubramanium} %
\affiliation{Department of Physical and Environmental Sciences,
  University of Toronto Scarborough, Toronto, Ontario, M1C 1A4,
  Canada} %
\author{Ilya G. Ryabinkin} %
\affiliation{Department of Physical and Environmental Sciences,
  University of Toronto Scarborough, Toronto, Ontario, M1C 1A4,
  Canada} %
\affiliation{Chemical Physics Theory Group, Department of Chemistry,
  University of Toronto, Toronto, Ontario M5S 3H6, Canada} %
\author{Artur F. Izmaylov} %
\affiliation{Department of Physical and Environmental Sciences,
  University of Toronto Scarborough, Toronto, Ontario, M1C 1A4,
  Canada} %
\affiliation{Chemical Physics Theory Group, Department of Chemistry,
  University of Toronto, Toronto, Ontario M5S 3H6, Canada} %

\date{\today}

\begin{abstract}
On-the-fly quantum nonadiabatic dynamics for large 
systems greatly benefits from the adiabatic representation readily available from the electronic structure 
programs. However, frequently occurring in this representation 
conical intersections introduce non-trivial geometric or Berry phases which require a special treatment 
for adequate modelling of the nuclear dynamics. 
We analyze two approaches for nonadiabatic dynamics using the
time-dependent variational principle and the adiabatic representation. 
The first approach employs adiabatic electronic functions with global
parametric dependence on the nuclear coordinates. The second approach uses 
adiabatic electronic functions obtained only at the centres of moving localized  
nuclear basis functions (e.g. frozen-width Gaussians).
Unless a gauge transformation is used to enforce single-valued boundary conditions,
the first approach fails to capture the geometric phase.
In contrast, the second approach accounts for the geometric phase naturally 
because of the absence of the global nuclear coordinate dependence in the electronic functions. 
\end{abstract}


\maketitle



The time-dependent variational principle (TDVP)\cite{Book/Kramer:1981,Dirac:1930/PCPS/376,Book/Frenkel:1934} provides a very efficient framework for
simulating quantum dynamics in large molecular systems. The most powerful aspect of this   
framework is use of time-dependent basis functions which reduces basis set 
size requirement compare to that for static basis sets. Two most widely used branches of the TDVP 
methodology constitute approaches related to the multi-configuration time-dependent 
Hartree (MCTDH) method\cite{mey90:73,Wang:2003/jcp/1289,mctdh} 
and approaches using frozen-width Gaussian functions.
\cite{Yang:2009ja,BenNun:2002tx,Shalashilin:2009/JCP/244101,Burghardt:2008iz,Worth:2008/MP/2077,Worth:2004/FD/307,Izmaylov:2013fe}
If the MCTDH-based approaches are more suitable for fixed diabatic models, 
the frozen Gaussian functions have been extended to simulating nuclear dynamics
with the on-the-fly calculation of the electronic potential energy surfaces.
\cite{Yang:2009ja,BenNun:2002tx,Shalashilin:2009/JCP/244101,Saita:2012di} 
Naturally, the adiabatic representation becomes the most straightforward representation 
for the electronic part of the problem in this case.

One of the most frequent manifestations of the nuclear quantum character is 
nonadiabatic phenomena where the nuclear dynamics involves several electronic states.
TDVP has been successfully extended and applied to modelling nonadiabatic dynamics (NAD). 
Very frequently NAD becomes necessary because adiabatic electronic potential energy 
surfaces form conical intersections (CIs).\cite{Domcke:2012/arpc/325,Yarkony:1996/rmp/985,
Yarkony:1998/acr/511,Yarkony:2001/jpca/6277}
CIs promote transitions between electronic states and introduce nontrivial geometric 
phases\cite{LonguetHigg:1958/rspa/1,Mead:1979/jcp/2284,Berry:1984/rspa/45,Mead:1992/rmp/51,Wittig:2012/pccp/6409,Althorpe:2006/jcp/084105} 
that can affect 
dynamics in profound ways.\cite{Kendrick:2003/jpca/6739,Ryabinkin:2013/prl/220406,Loic:2013/jcp/234103,Ryabinkin:2014/jcp/214116,Guo:2016/jacs,Hazra:2015he} 
It is important to stress that CIs and associated GPs
appear only when one uses the adiabatic representation for description of 
electronic part of the total wave-function. CIs and GPs disappear when the 
diabatic representation is used, however, 
physical observables of NAD do not change with the representation. Therefore,
the dynamical features that emerge in the adiabatic representation due to a nontrivial 
GP appear in the diabatic or any other representation as 
well.\cite{PhysRevLett.113.263004,PhysRevA.93.042108}  

\begin{figure}
  \includegraphics[width=0.4\textwidth]{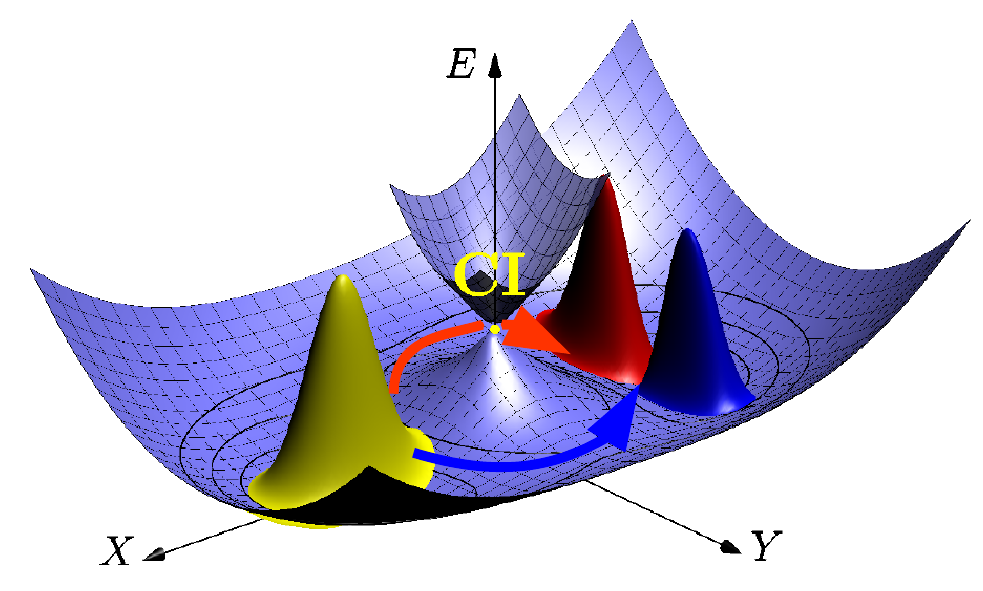}
  \caption{Destructive interference due to geometric phase in low
    energy dynamics: the initial nuclear density is in yellow and 
    the one at a later time is in red-blue.}
  \label{fig:GPdi}
\end{figure}

One of the simplest signatures of the nontrivial GP introduced by CI is a nodal line 
appearing in non-stationary nuclear density that moves between minima of a double-well 
potential with a CI in between the minima (\fig{fig:GPdi}). This nodal line appears due to acquisition of 
opposite GPs by parts of the wave-packet going around the CI from different sides.
\cite{Schon:1995/jcp/9292,Ryabinkin:2013/prl/220406,Guo:2016/jacs} 
This destructive interference can significantly slow down the transfer between the minima and
even freeze it completely.\cite{Ryabinkin:2013/prl/220406} 
Of course, any accurate method of quantum dynamics 
must reproduce the nodal line appearing in this setup. 
Mead and Truhlar\cite{Mead:1979/jcp/2284} shown that 
to capture the GP in simulations using time-independent nuclear basis functions 
it is necessary to introduce a complex-valued gauge transformation. This transformation 
requires some global information about the topology of the potential energy surfaces forming CI 
and thus poses difficulties in application within the on-the-fly framework, 
where only local information is available.   
To address this difficulty, we consider two approaches to formulating 
TDVP using the adiabatic representation and show how the GP can be accounted 
in each of these approaches.


The total non-relativistic molecular Hamiltonian can be written as 
\bea
\hat H(\rr,\RR) = \hat T_N + \hat H_e(\rr;\RR), 
\eea
where $\hat T_N = -\nabla_{\RR}^2/2$ is the kinetic energy of nuclei,
\footnote{For simplicity, we use the nuclear coordinates in the mass-weighted form, 
and atomic units are used throughout this paper.}  and 
$\hat H_e(\rr;\RR)$ is the electronic Hamiltonian with electronic $\rr$ and nuclear $\RR$ 
coordinates. $\hat H_e(\rr;\RR)$ determines the adiabatic electronic wave-functions 
$\ket{\phi_{s}(\RR)}$ and potential energy surfaces $E_e^{(s)}(\RR)$:
$\hat H_e(\rr;\RR)\langle \rr\ket{\phi_{s}(\RR)} = E_e^{(s)}(\RR)\langle \rr\ket{\phi_{s}(\RR)}$.

\paragraph{Global adiabatic (GA) representation:}
The total non-stationary wave-function can be expanded in the 
adiabatic representation as
\bea\label{eq:WFad1}
\langle\rr,\RR\ket{\Psi(t)} = \sum_{I,s} C_{I}^{(s)}(t) \langle \RR \ket{G_{I}^{(s)}}
\langle\rr\ket{\phi_{s}(\RR)},
\eea
where  $C_{I}^{(s)}$ are time dependent coefficients, indices $s$ and $I$ enumerate
the electronic and nuclear coherent states (CSs)  
\bea\notag
\langle \RR \ket{G_{I}^{(s)}} &=& \prod_{j=1}^{N} \left(\frac{\omega_j}{\pi}\right)^{1/4} 
\exp\Big{[} -\frac{\omega_j}{2}[R_j -q_{jI}^{(s)}(t)]^2  \\ \label{eq:CS}
&&+ip_{jI}^{(s)}[R_j -q_{jI}^{(s)}(t)]+\frac{i}{2} p_{jI}^{(s)}q_{jI}^{(s)} \Big{]}
\eea
 with time-dependent positions $\qq_I^{(s)} = \{q_{jI}^{(s)}\}_{j=1,N}$ and 
 momenta $\pp_I^{(s)}=\{p_{jI}^{(s)}\}_{j=1,N}$ [$N={\rm dim}(\RR)$].

Equations of motion (EOM) for positions and momenta of CSs can be obtained using 
TDVP but resulting EOM would introduce unnecessary complexity for our consideration. 
Thus, here, we adopt simpler EOM that follow classical dynamics on the adiabatic potential 
energy surfaces
\bea\label{eq:qdot}
\dot{\qq}_{I}^{(s)} &=& \pp_{I}^{(s)} \\ \label{eq:pdot}
\dot{\pp}_{I}^{(s)} &=& -\frac{E_{e}^{(s)}(\RR)}{\partial \RR}\Big{\vert}_{\RR = \qq_{I}^{(s)}}. 
\eea
This simplifies variation of the total wave-function by restricting it only to the linear 
coefficients $C_{I}^{(s)}(t)$ 
\bea\label{eq:var1}
\delta\langle\rr,\RR \ket{\Psi(t)} = \sum_{I,s} [\delta C_I^{(s)}(t)] \langle\RR \ket{G_{I}^{(s)}}\langle\rr\ket{\phi_{s}(\RR)}.
\eea
 Applying the Dirac-Frenkel 
 TDVP\cite{Dirac:1930/PCPS/376,Book/Frenkel:1934} 
 \bea
 \bra{\delta \Psi} \hat H-i\partial_t \ket{\Psi} = 0
 \eea
and substituting the $\Psi$ and $\delta \Psi$ expressions from Eqs.~\eqref{eq:WFad1} and \eqref{eq:var1}
we obtain
\bea
\sum_{IK,ss'} \delta C_{I}^{(s)} \bra{G_{I}^{(s)} } \hat H_{N}^{(ss')} - i\partial_t \ket{G_{K}^{(s')}}C_{K}^{(s')} = 0,
\eea
where 
\bea
\hat H_{N}^{(ss')} &=& \bra{\phi_s(\RR)} \hat H_e(\RR) + \hat T_N \ket{\phi_{s'}(\RR)} \\
&=& E_e^{(s)}(\RR)\delta_{ss'} + \hat T_N + \hat \tau_{ss'}, \\ \notag
\hat \tau_{ss'} &=& -\bra{\phi_s(\RR)} \nabla_{\RR} \phi_{s'}(\RR)\rangle \nabla_{\RR} \\ \label{eq:NACs}
&& - \bra{\phi_s(\RR)} \nabla_{\RR}^2 \phi_{s'}(\RR)\rangle/2. 
\eea
Note that so-called nonadiabatic couplings (NACs) $\hat \tau_{ss'}$ appear as a 
result of a global dependence of the electronic wave-functions $\phi_{s}$
on the nuclear coordinates $\RR$. 
Considering independence of $\delta C_{I}^{(s)}$ variations, EOM for the coefficients $C_{I}^{(s)}(t)$
become
\bea
\sum_{K,s'} \bra{G_{I}^{(s)}} \hat H_{N}^{(ss')} - i\partial_t \ket{G_{K}^{(s')}}C_{K}^{(s')} = 0,
\eea
Rearranging few terms leads to EOM in the form
\bea \notag
i\dot{C}_J^{(s)} &=& \sum_{I,K}\bra{G_{J}^{(s)}}G_{I}^{(s)}\rangle^{-1}
\Big{[}\bra{G_{I}^{(s)}} \hat H_{N}^{(ss')}\ket{G_{K}^{(s')}}\\
\label{eq:GAeom}
&&-i\bra{G_{I}^{(s)}}\partial_t G_{K}^{(s')}\rangle\Big{]}C_{K}^{(s')}.
\eea
where $\bra{G_{J}^{(s)}}G_{I}^{(s)}\rangle^{-1}$ are elements of the inverse CS overlap matrix. 
Time-derivatives of CSs needed in \eq{eq:GAeom} are
derived using the chain rule
\bea
\ket{\partial_t G_{K}^{(s')}} &=&  \ket{\frac{\partial G_{K}^{(s')}}{\partial \qq_{K}^{(s')}}} 
\dot{\qq}_{K}^{(s')}(t) + \ket{\frac{\partial G_{K}^{(s')}}{\partial \pp_{K}^{(s')}}}\dot{\pp}_{K}^{(s')}(t).
\eea

The difficulty associated with a proper treatment of the nuclear dynamics using 
global adiabatic electronic functions is that $\ket{\phi_s(\RR)}$ are 
double-valued functions with respect to $\RR$ in the CI case. 
To have a single-valued total wave-function in \eq{eq:WFad1} 
the nuclear wave-function must also be double-valued, which is not the case for typical Gaussian-like 
basis sets [\eq{eq:CS}]. In order to include GP related effects in the nuclear dynamics 
one needs to substitute the real but double-valued adiabatic electronic wave-functions $\ket{\phi_s(\RR)}$
by their complex but single-valued counterparts: 
$\ket{\tilde{\phi}_s(\RR)} = e^{i\theta_s(\RR)}\ket{\phi_s(\RR)}$, where $e^{i\theta_s(\RR)}$ 
is a phase factor that changes its sign when $\RR$ follows any curve encircling the CI. 
This phase factor can be seen as a gauge transformation which is needed  
when a single-valued basis functions for the nuclear counterpart are used.   

\paragraph{Moving crude adiabatic (MCA) representation:}
Alternatively, EOM can be derived using a different ansatz for the total wave-function 
\bea\label{eq:WF2}
\langle\rr,\RR\ket{\Psi(t)} = \sum_{I,s} C_{I}^{(s)}(t) \langle\RR \ket{G_{I}^{(s)}}
\langle\rr\ket{\phi_{s}(\qq_I^{(s)})},
\eea
here the electronic functions are evaluated only at the centres of CSs, $\qq_I^{(s)}$, and thus 
do not depend on the nuclear coordinates $\RR$. To simplify the notation 
we will denote $\ket{\phi_{s}(\qq_I^{(s)})}$ as $\ket{\phi_{I}^{(s)}}$.
Treating CS motion classically [Eqs.~\eqref{eq:qdot} and \eqref{eq:pdot}] 
we repeat the derivation of EOM for $C_{I}^{(s)}(t)$ in \eq{eq:WF2} and obtain 
\bea \notag
i\dot{C}_{J}^{(s'')} &=& \sum_{I,K} \bra{G_{J}^{(s'')}\phi_{J}^{(s'')}}G_{I}^{(s)}\phi_{I}^{(s)}\rangle^{-1} \Big{[}\bra{G_{I}^{(s)}} \hat H_{IK}^{(ss')} \ket{G_{K}^{(s')}}\\ \label{eq:TDCAeom}
&&-i\bra{\phi_{I}^{(s)}G_{I}^{(s)}}\phi_{K}^{(s')}\partial_t G_{K}^{(s')}\rangle\Big{]}C_{K}^{(s')},
\eea
 where $\bra{G_{J}^{(s'')}\phi_{J}^{(s'')}}G_{I}^{(s)}\phi_{I}^{(s)}\rangle^{-1}$ 
 are elements of the total inverse overlap matrix, and 
 \bea\label{eq:Hcad}
 \hat H_{IK}^{(ss')} &=& \bra{\phi_{I}^{(s)}} \hat H_e(\RR) \ket{\phi_{K}^{(s')}} + 
 \bra{\phi_{I}^{(s)}}\phi_{K}^{(s')}\rangle \hat T_N \\ \notag 
 &&-i  \bra{\phi_{I}^{(s)}}\partial_t \phi_{K}^{(s')}\rangle.
 \eea
Here, the adiabatic electronic functions obtained at different points of nuclear geometry
and corresponding to different electronic states are non-orthogonal: 
$\bra{\phi_I^{(s)}}\phi_K^{(s')}\rangle\ne\delta_{ss'}$ if $I\ne K$. Also, the electronic
functions are not eigenfunctions of the electronic Hamiltonian for all values of $\RR$,
therefore, $\bra{\phi_{I}^{(s)}} \hat H_e(\RR) \ket{\phi_{K}^{(s')}}$ is a 
$\RR$ and $t$ dependent matrix of functions 


Using the chain rule, the electronic time-derivative couplings in \eq{eq:Hcad} 
can be expressed as 
\bea\label{eq:MCAder}
\bra{\phi_{I}^{(s)}}\partial_t \phi_{K}^{(s')}\rangle &=& 
\bra{\phi_{I}^{(s)}}\frac{\partial\phi_{K}^{(s')}}{\partial \qq_{K}^{(s')}}\rangle \dot{\qq}_{K}^{(s')} . 
\eea
Considering the equivalence between dependencies of the MCA electronic functions on 
centres of CSs and the GA electronic functions on $\RR$, the electronic function 
derivatives in \eq{eq:MCAder} 
are similar to the first order derivative part of NACs in \eq{eq:NACs}. 
The first order derivative couplings diverge at the point of the CI, 
however, since the CS centres form a measure zero subset, CSs will never have their centres 
exactly at the CI seam. Note that in the MCA representation there are no analogues of the
second order derivative parts of NACs. The second order derivatives in NACs pose difficulties for 
integrating EOM due to their $1/R^2$ divergent behavior with the distance
from the CI $R$.\cite{meek:2016c} 

From the GP point of view, the MCA formalism can be thought as
a truly diabatic formalism since the electronic functions do not have the dependence on $\RR$, 
and thus problems emerging in the GA representation do not appear here. 
Nevertheless, due to a parametric dependence of the adiabatic electronic functions
on CSs' centres, the MCA representation has GPs carried by the electronic functions.  
   


We illustrate nuclear dynamics in the introduced representations for a 2D-LVC model where 
formulated EOM can be simulated without additional approximations and where the GP plays 
a significant role. The total Hamiltonian for 2D-LVC is 
\begin{equation}
  \label{eq:H_lvc}
 \hat H_{\rm LVC} = 
  \begin{pmatrix} 
    \hat T_N + V_{11} & V_{12} \\
    V_{12} & \hat T_N + V_{22}
  \end{pmatrix},
\end{equation}
where $\hat T_N = -\frac{1}{2}(\partial^2
/\partial x^2 +\partial^2 /\partial y^2) $ is the nuclear kinetic
energy operator, 
 $ V_{11}$ and
$ V_{22}$ are the diabatic potentials represented by identical 2D
parabolas shifted in the $x$-direction by $a$ 
\begin{align}
  \label{eq:diab-me-11}
  V_{11}(\RR) = {} & \frac{\omega^2}{2}\left[(x + a)^2 + y^2\right],\\
  \label{eq:diab-me-22}
  V_{22}(\RR) = {} & \frac{\omega^2}{2}\left[(x - a)^2 + y^2\right].
\end{align}
To have the CI in the adiabatic representation, $V_{11}$ and
$V_{22}$ are coupled by a linear potential $V_{12}(\RR)=c y$.
Thus for this example we have $\RR=(x,y)$ and 
the electronic Hamiltonian can be defined as 
$\hat H_e(\RR) = \sum_{ij} \ket{\varphi_i}V_{ij}(\RR)\bra{\varphi_j}$, where 
$\ket{\varphi_i}$'s are the diabatic electronic states. 

Switching to the adiabatic representation 
is done by rotating the electronic basis into the adiabatic states
\begin{eqnarray}\label{eq:adi1}
  \ket{\phi_1(\RR)} & = & \phantom{-}\cos\theta(\RR)\,\ket{\varphi_1} +
  \sin\theta(\RR)\,\ket{\varphi_2}, \\ \label{eq:adi2}
  \ket{\phi_2(\RR)} & = & -\sin\theta(\RR)\,\ket{\varphi_1} +
  \cos\theta(\RR)\,\ket{\varphi_2},
\end{eqnarray}
which diagonalize the potential matrix.
$\theta(\RR)$ is a rotation angle 
\begin{equation}
  \label{eq:theta}
  \theta = \frac{1}{2}\arctan \dfrac{2\,V_{12}}{V_{22} - V_{11}}. 
\end{equation}
If we track $\theta$ changes continuously along a contour encircling 
the CI, it will change by $\pi$, which flips the sign of the phase factor 
$e^{i\theta}$.\cite{Izmaylov:2016eo}
The nuclear 2D-LVC Hamiltonian in the adiabatic representation is 
\begin{equation}
  \label{eq:Had}
   \hat H_\text{adi}  =   
  \begin{pmatrix}
   \hat T_N + \hat\tau_{11}& \hat\tau_{12} \\
    \hat\tau_{21} & \hat T_N +\hat\tau_{22}
  \end{pmatrix} +
  \begin{pmatrix}
    E_e^{(-)} & 0 \\
    0 & E_e^{(+)}
  \end{pmatrix},
\end{equation}
where
\begin{align}
  \label{eq:Wmin}
  E_e^{(\pm)} = & {} \dfrac{1}{2}\left(V_{11} + V_{22}\right) \pm
  \dfrac{1}{2}\sqrt{\left(V_{11} - V_{22}\right)^2 + 4 V_{12}^2}
\end{align} 
are the adiabatic energy surfaces and
\begin{align}
  \label{eq:tau-adi-diag}
  \hat\tau_{11} & {} =\hat\tau_{22} = \frac{1}{2}
  \nabla\theta\cdot\nabla\theta  \\ \label{eq:tau-adi-offd}
  \hat\tau_{12} & {} = -\hat\tau_{21} =
  \frac{1}{2}\left(\nabla^2\theta+2\nabla\theta\cdot\nabla\right)
\end{align}
are NACs.

In order to include the GP we use the gauge transformation of the electronic functions 
that can be seen as a modification of the nuclear Hamiltonian
$ H_\text{adi}^{\rm GP} = e^{-i\theta}\hat H_{\rm adi} e^{i\theta}$.\cite{Ryabinkin:2014/jcp/214116} 
This transformation leads to modification of NACs
\begin{equation}
  \label{eq:HadGP}
  H_\text{adi}^{\rm GP} =   
  \begin{pmatrix}
    \hat T_N + \hat\tau_{11}^{\rm GP} & \hat\tau_{12}^{\rm GP} \\
    \hat\tau_{21}^{\rm GP} & \hat T_N + \hat\tau_{22}^{\rm GP}
  \end{pmatrix} +
  \begin{pmatrix}
    E_e^{(-)} & 0 \\
    0 & E_e^{(+)}
  \end{pmatrix},
\end{equation}
where 
\bea\label{eq:tauGP11}
\hat \tau_{11}^{\rm GP} = \hat \tau_{22}^{\rm GP} &=& (\nabla\theta)^2-\frac{i}{2}(\nabla^2\theta+2\nabla\theta\nabla),\\
\hat \tau_{12}^{\rm GP}= -\hat\tau_{21}^{\rm GP} &=& -i(\nabla\theta)^2+\frac{1}{2}(\nabla^2\theta+2\nabla\theta\nabla).
\eea

For the MCA representation, the electronic states are calculated as
\begin{eqnarray}\label{eq:cadi}
  \ket{\phi_{I}^{(1)}} & = & \phantom{-}\cos\theta(\qq_{I}^{(1)})\,\ket{\varphi_1} +
  \sin\theta(\qq_{I}^{(1)})\,\ket{\varphi_2}, \\ \label{eq:adi2}
  \ket{\phi_{I}^{(2)}} & = & -\sin\theta(\qq_{I}^{(2)})\,\ket{\varphi_1} +
  \cos\theta(\qq_{I}^{(2)})\,\ket{\varphi_2},
\end{eqnarray}
where $q_I^{(1)}$ and $q_I^{(2)}$ are centres of corresponding CSs.
Therefore, integrals $\bra{\phi_{I}^{(s)}}H_e(\RR)\ket{\phi_J^{(s)}}$ for the 2D-LVC 
model are simply linear combinations of $V_{kl}(\RR)$ multiplied by 
$\cos$ and $\sin$ functions.  

\begin{figure}
  \centering
  \includegraphics[width=0.5\textwidth]{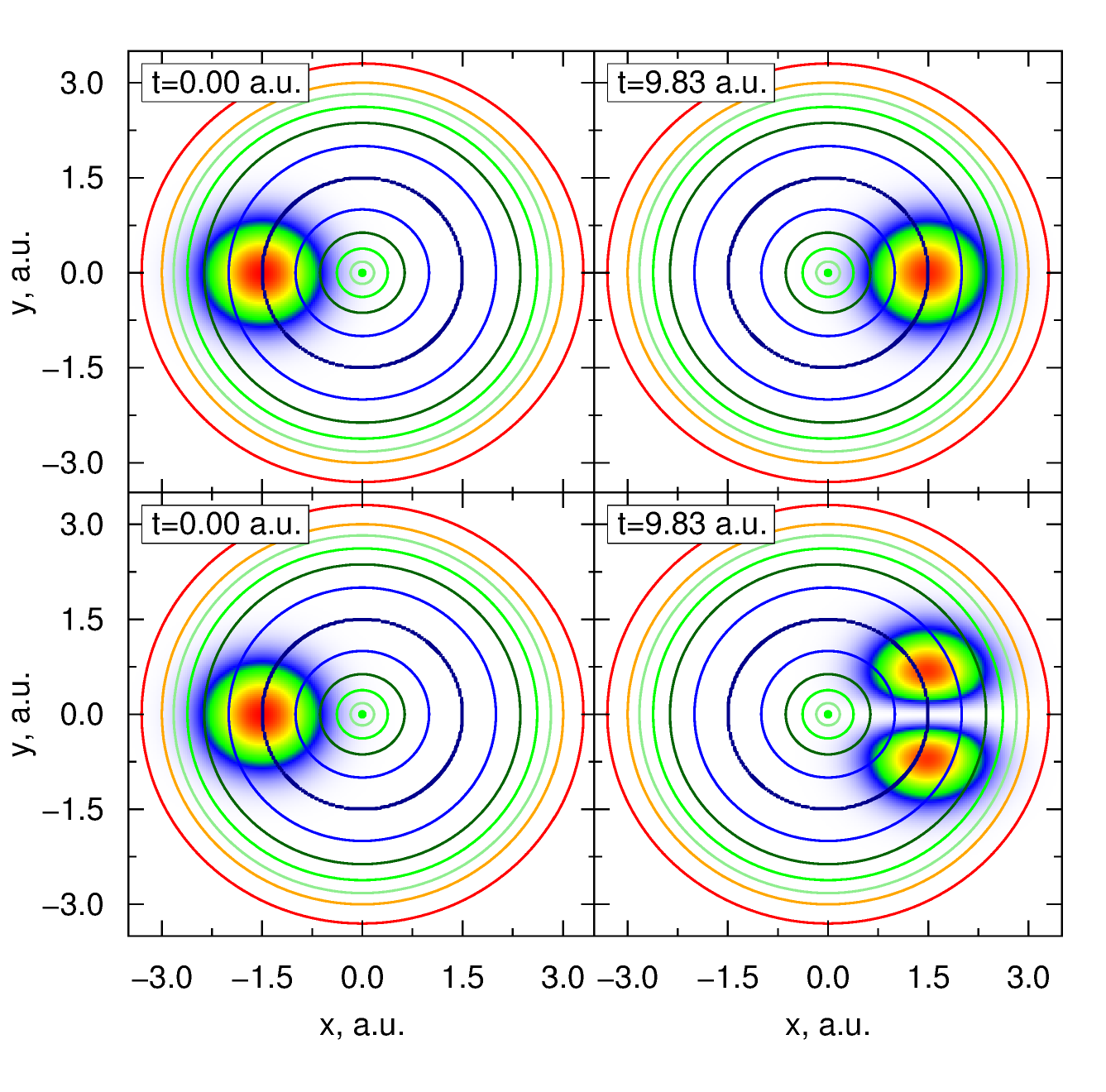}
  \caption{Nuclear probability densities for the approach ignoring GP (top) and  including GP (bottom): the initial density distributions are on the left panels, density distributions at a later time on the right panels. 
Contours of the ground adiabatic state potential from $H_{\rm LVC}$ ($\omega=2$, $a=1.5$, $c=6$) 
are superimposed on the density distributions.}
  \label{fig:node}
\end{figure}

To illustrate the performance of all three approaches in reproducing the GP
we simulate nuclear dynamics of the initial wave-function 
\bea
\langle \RR \ket{\Psi(t=0)} = \frac{\ket{\phi_1}}{\sqrt{2}} [\langle \RR \ket{g_1^{(1)}}+
\langle \RR \ket{g_2^{(1)}}]
\eea
that is comprised of two CSs, $\ket{G_{I}^{(1)}} = \ket{G_{I}^{(1)}(q,p_I)}$
centred at the same point $q=(-1.5,0)$ of the ground potential energy surface, 
but with momenta $p_1=(0.1,0.5)$ and $p_2=(0.1,-0.5)$, which have the 
opposite $y$-components. Using three different Hamiltonians, \eqs{eq:Hcad}, \eqref{eq:Had}, and 
\eqref{eq:HadGP}, we simulate time-dependent wave-functions and monitor 
the total nuclear density $\rho_n(\RR,t) = {\rm Tr}_e[\langle\RR\ket{\Psi(t)}\bra{\Psi(t)}\RR\rangle]$, 
where ${\rm Tr}_e$ is the trace over the electronic coordinates. Figure~\ref{fig:node} 
illustrates that dynamics with the adiabatic Hamiltonian \eqref{eq:Had} misses the GP, 
while two other Hamiltonians reproduce the GP induced destructive interference perfectly.
However, mechanisms for the destructive interference in the two approaches is quite 
different: For the GA representation, two CSs acquire different phases from the 
$-i\nabla\theta\nabla$ part of the diagonal NAC [\eq{eq:tauGP11}] because 
the $\theta$ angle is proportional to the geometric angle between the initial and final positions of a
CS with respect to the CI. In the MCA representation, CSs acquire different phases 
due to GPs of the associated electronic wave-functions.    

In conclusion, we illustrated that GP effects can be successfully captured 
using both global and moving crude
adiabatic representations. However, capturing the GP in the GA representation  
seems difficult for calculations beyond models. The systematic application of TDVP 
with the GA electronic functions becomes especially difficult for 
the on-the-fly calculations, not only because of the necessity to generate 
the gauge transformation $e^{i\theta(\RR)}$ using only local information but also because 
of the second order derivative NACs whose integrals with Gaussians are divergent.\cite{meek:2016c} 
On both accounts, employing the MCA representation is much more practical: 
GPs are always carried by the electronic functions and numerically difficult
second order derivative NACs never appear in the formalism. 
Interestingly, in previous works on ab initio multiple spawning (AIMS),
\cite{Martinez:1996ti,ben-nun:7244,Martinez2000rev,Yang:2009ja,BenNun:2002tx} 
the derivation was presented starting with the 
GA representation but the actual working EOM for the linear coefficients 
were very similar to the ones obtained in the current work using the MCA representation. 
This was the result of approximations needed to make AIMS EOM feasible for simulating 
dynamics in realistic systems. The current work provides a rigorous framework of the 
MCA representation that justifies some of the approximations made in AIMS. 
Also, the MCA representation can be seen as an effortless realization of a recently 
proposed on-the-fly diabatization to solve the problem of numerical difficulties in integration 
of the second order NACs.\cite{meek:2016d}  

{\it Acknowledgments: }
Authors are grateful to Todd Martinez, Benjamin Levine, and Michael Schuurman 
for stimulating discussions. A.F.I. acknowledges funding from a Sloan Research Fellowship 
and the Natural Sciences and Engineering Research Council  
of Canada (NSERC) through the Discovery Grants Program.


%

\end{document}